\documentclass[aps,twocolumn,nofootinbib,preprintnumbers,superscriptaddress,eqsecnum]{revtex4-1}

\usepackage[utf8]{inputenc}
\usepackage{color}
\usepackage[normalem]{ulem}
\usepackage{amsmath}
\usepackage{enumerate}
\usepackage{amsfonts}
\usepackage{epsfig}
\usepackage[
      colorlinks=true,
      linkcolor=blue,
      urlcolor=blue,
      filecolor=black,
      citecolor=red,
      pdfstartview=FitV,
      pdftitle={},
        pdfauthor={},
        pdfsubject={},
        pdfkeywords={},
        pdfpagemode=None,
        bookmarksopen=true
      ]{hyperref}
\usepackage{subcaption}      
      
\newcommand{\be}{\begin{equation}}
\newcommand{\ee}{\end{equation}}
\newcommand{\bea}{\begin{eqnarray}}
\newcommand{\eea}{\end{eqnarray}}

\newcommand{\bln}{\begin{align}}
\newcommand{\eln}{\end{align}}
\newcommand{\bst}{\begin{split}}
\newcommand{\est}{\end{split}}
\newcommand{\bi}{\begin{itemize}}
\newcommand{\ei}{\end{itemize}}
\newcommand{\ben}{\begin{enumerate}}
\newcommand{\een}{\end{enumerate}}

\def\le{\left}
\def\ri{\right}
\def\ha{{1\over 2}}

\def\Lam{{\Lambda}}

\def\al{{\alpha}}

\def\th{{\theta}}

\def \th{{\theta}}

\def\sig{{\sigma}}
\def\ep{{\varepsilon}}

\newcommand{\p}{\partial}

\def\eeq{\end{equation}}

\newcommand\sM{{\ensuremath{{\mathcal M}}}}

\newcommand\sO{{\ensuremath{{\mathcal O}}}}

\begin{document}

\title {Goldstone modes and photonization for higher form symmetries}

\author{Diego M. Hofman}
\email{d.m.hofman@uva.nl}
\affiliation{Institute for Theoretical Physics, University of Amsterdam, Science Park 904, Postbus 94485, 1090 GL Amsterdam, The Netherlands}
\author{Nabil Iqbal}
\email{nabil.iqbal@durham.ac.uk}
\affiliation{Centre for Particle Theory, Department of Mathematical Sciences, Durham University,
South Road, Durham DH1 3LE, UK}

\begin{abstract}
We discuss generalized global symmetries and their breaking. We extend Goldstone's theorem to higher form symmetries by showing that a perimeter law for an extended $p$-dimensional defect operator charged under a continuous $p$-form generalized global symmetry necessarily results in a gapless mode in the spectrum. We also show that a $p$-form symmetry in a conformal theory in $2(p+1)$ dimensions has a free realization. In four dimensions this means any 1-form symmetry in a $CFT_4$ can be realized by free Maxwell electrodynamics, i.e. the current can be {\it photonized}. The theory has infinitely many conserved 0-form charges that are constructed by integrating the symmetry currents against suitable 1-forms. We study these charges by developing a twistor-based formalism that is a 4d analogue of the usual holomorphic complex analysis familiar in $CFT_2$. The charges are shown to obey an algebra with central extension, which is an analogue of the 2d Abelian Kac-Moody algebra for higher form symmetries. 
\end{abstract}

\maketitle

\newpage
\begingroup
\hypersetup{linkcolor=black}
\endgroup

\section{Introduction}

In this short note we discuss aspects of generalized global symmetries \cite{Gaiotto:2014kfa}. A $p$-form generalized global symmetry is an invariance of a theory parametrized by a closed $p$-form, resulting (in the continuous case) in a divergenceless $p+1$-form current $J$:
\be
d \star J = 0\, .
\ee  
The $0$-form case is just a conventional global symmetry and results in a conserved particle number, whereas the higher-form generalizations result in a conserved density of higher dimensional objects (strings, branes, etc.). Thus the operator charged under a $p$-form symmetry is $p$-dimensional: in the conventional 0-form case we have a point-like local operator that creates a particle, in the $1$-form case we have a line-like operator (e.g. a Wilson or t'Hooft line \cite{Kapustin:2005py,Aharony:2013hda}) that creates a string, etc. 

Though they may sound unfamiliar, such generalized symmetries are in fact present in many very familiar theories. They have recently found applications in diverse contexts, ranging from constraining the phase structure of gauge theories and topological phases \cite{Gaiotto:2017yup,Komargodski:2017dmc,PhysRevB.93.155131,Wang:2014pma,Tanizaki:2017mtm,Kitano:2017jng,Gaiotto:2017tne} to a new symmetry-based formulation of magnetohydrodynamics \cite{Grozdanov:2016tdf}. 

At the heart of these applications lies the fact that higher-form symmetries are just as powerful as conventional global symmetries: they may be coupled to external sources, they sometimes have anomalies and they can spontaneously break. Following Landau, we should then classify low-energy phases of matter by studying the realization of these symmetries. An order parameter for the potential breaking of the symmetry is the long-distance behavior of the charged operator. For a $p$-form symmetry, a charged object $W(C)$ is defined on a $p$-dimensional submanifold $C$. 

There are two possible phases: in the first, the symmetry is {\it unbroken}. In this case, we find at long distances
\be
\langle W(C) \rangle \sim e^{-T_{p+1} \mathrm{Area}(C)} \label{area} \, ,
\ee
where Area$(C)$ denotes the volume of a (minimal) $p+1$ dimensional manifold $B$ that ``fills in'' $C$ such that $\p B = C$ and $T_{p+1}$ should be understood as the non-zero tension of the $p+1$ dimensional objects that $J$ counts. This is the analogue of an exponentially decaying correlation function $\langle \sO^{\dagger}(x) \sO(0) \rangle \sim \exp(-m |x|)$ in the $0$-form case, and in the 1-form case it is literally an area law for the line-like operator. If $C$ is topologically non-trivial, then there exists no $B$ and $W(C)$ vanishes. 

In the other phase, the symmetry is {\it spontaneously broken}. In this case, we find
\be
\langle W(C) \rangle \sim  e^{-T_{p} \mathrm{Perimeter}(C)} \, , \label{perimeter} 
\ee
where Perimeter$(C)$ is understood to denote the $p$-volume of the $p$-dimensional manifold $C$ itself.\footnote{This is somewhat imprecise: it may actually include any {\it local} functional of the geometry of $C$, e.g. an extrinsic curvature, etc.} This only depends locally on $C$ and so is the analogue of a factorized correlation function $\langle \sO^{\dagger}(x) \sO(0) \rangle \sim \langle \sO^{\dagger} \rangle \langle \sO \rangle$ in the $0$-form case. 

\section{Higher form Goldstone modes}
\label{sec:goldstone}
We now present a higher-form generalization of Goldstone's theorem to show that a spontaneously broken $p$-form symmetry results in a gapless Goldstone mode. The Ward identity for $J$ in the presence of $W(C)$ is \cite{Gaiotto:2014kfa}:
\be
(d\star J(x)) W(C) = i q \delta_C(x) W(C) \label{ward}\, ,
\ee
where $q$ is the charge of $W(C)$ and $\delta_C(x)$ is a $(d-p)$-form delta function\footnote{More precisely it is a $d-p$ form that is zero away from $C$ and such that for any $p$-form field $B_p$ we have
$$
\int_{\mathbb{R}^d} B_p \wedge \delta_C(x) = \int_C B_p \ . 
$$}
with support on $C$. 

Now, suppose that we are in a spontaneously broken phase. As $W(C)$ obeys the perimeter law \eqref{perimeter}, we may define a new renormalized $\overline{W}(C)$ by stripping off the perimeter dependence:
\be
\overline{W}(C) \equiv W(C) e^{+T_p \mathrm{Perimeter}(C)} \ . 
\ee 
This satisfies the same Ward identity \eqref{ward} and is still defined locally on $C$. 

\begin{figure}
\begin{center}
\includegraphics[scale=0.4]{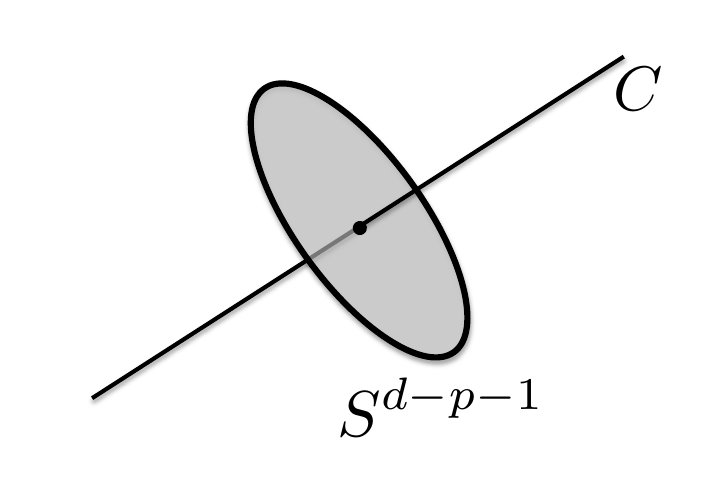}
\end{center}
\caption{Geometry for Goldstone theorem: $C$ is an infinite $p$-plane that intersects $B_{d-p}$ at a single point; $S^{d-p-1}$ is the boundary of $B_{d-p}$.} 
\label{fig:goldstone}
\end{figure}

Let us now take $C$ to be an infinite flat $p$-plane and consider a $(d-p)$ dimensional ball $B_{d-p}$ with radius $R$ that intersects $C$ at a single point, as shown in Figure \ref{fig:goldstone}. The boundary of $B_{d-p}$ is a $S^{d-p-1}$ that wraps $C$, and $R$ is the perpendicular distance from $C$ to the $S^{d-p-1}$. Integrating both sides of \eqref{ward} over this ball and using Gauss's law we find
\be
\overline{W}(C) \int_{S^{d-p-1}(R)} \star J = i\, q \, \overline{W}(C) \, . \label{maineq}
\ee 

Now take the vacuum expectation value of both sides of this expression. As we have removed the leading perimeter dependence, by construction $\overline{W}(C)$ will have a finite expectation value: $\overline{W}(C) = c$. Now as we increase $R$ the right-hand side and thus the left-hand side cannot change. But this means that the correlation function
\be
\langle J(R) \overline{W}(C) \rangle \sim \frac{i \,q\, c}{R^{d-p-1}}  \, ,\label{scaling}
\ee
where $J(R)$ is the current evaluated at a typical point on the $S^{d-p-1}$. There is thus a power-law correlation between the current and $\overline{W}(C)$. This implies that there is a massless excitation in the spectrum, which we will call the higher-form Goldstone mode. 

In particular, in the case that $p=0$, $C$ is just a single point and the argument is merely a reformulation of the usual Goldstone theorem in the language of Euclidean path integrals rather than commutators.

Let us now apply this machinery to Maxwell electrodynamics in four dimensions. In the absence of dynamical magnetic monopoles, this theory has a $p=1$ generalized global symmetry, with a conserved current that counts magnetic flux:
\be
J^{\mu\nu} = \ha \ep^{\mu\nu\rho\sig} F_{\rho\sig}, \label{Jdef}
\ee
where $F$ is the field strength. The line-like operator $W(C)$ charged under $J$ is the t'Hooft line \cite{Kapustin:2005py}; this corresponds to the insertion of an external magnetic monopole along a fixed worldline $C$. 

We now discuss the phases of electromagnetism from this point of view. Consider first free electrodynamics with no charged matter. In this case it is straightforward to show that $W(C)$ has a perimeter law, and thus the 1-form symmetry associated with \eqref{Jdef} is \textit{spontaneously broken}. The Goldstone mode is the usual {\it photon} \cite{Gaiotto:2014kfa}. To make this comparison clear we write the low-energy action and current as
\be
S = -\frac{g^2}{4} \int d^4x\;(d\tilde{A} - b)^2 \, ,\qquad J = g^2 \le( d\tilde{A} - b\ri) \, ,\label{dualac}
\ee
where $\tilde{A}$ is a 1-form that is conventionally called the {\it magnetic} photon, $b$ is an external 2-form source that couples to this current, and \eqref{dualac} is the electric-magnetic dual of the usual Maxwell action. The t'Hooft line is realized in terms of this low-energy field as $W(C) = \exp\le(i q\int_{C} \tilde{A}\ri)$. We see that \eqref{dualac} should actually be viewed as the Goldstone action for a spontaneously broken 1-form symmetry, where the usual $U(1)$ gauge coupling $g^2$ plays the role of the stiffness of the symmetry breaking. 


This description applies to our own universe provided we are at scales longer than the mass of the electron. Thus the fundamental (gauge invariant) principle protecting the masslessness of our photon is a spontaneously broken 1-form symmetry. The basic physical idea that condensation of extended objects leads to gapless modes is central to the idea of ``string-net condensation'' used to construct emergent photons from lattice models \cite{Levin:2004mi}, and our proof shows how such ideas may be understood in the continuum.  

In the Higgs (i.e. superconducting) phase of electromagnetism the behavior of magnetic flux is different: it is instead collimated into a tensionful Abrikosov-Nielsen-Olesen vortex. Thus $W(C)$ has an area law \eqref{area} with $T_{p+1}$ the tension of the vortex. The 1-form symmetry in this phase is {\it unbroken}. This provides an order parameter for superconductivity: it is well-known that there does not exist a conventional gauge-invariant order parameter using only local operators \cite{hansson2004superconductors}. See also \cite{Rosenstein:1990py, Kovner:1990pz}.

Returning to the general case, we see that for a spontaneously-broken $p$-form symmetry, the current $J$ is always realized nonlinearly in terms of a Goldstone $p$-form gauge field $B$ as $J = v^2 dB$ with action
\be\label{stiff}
S = -v^2 \int d^dx \, (dB)^2  \, ,
\ee 
where $v$ is a stiffness and where the coupling to the charged operator is $W(C) = \exp\le(iq \int_{C} B\ri)$. This construction saturates the broken Ward identity \eqref{ward}. 

\section{Conformality and photonization}
\label{sec:photonization}

We will now discuss a property of $p$-form symmetries in even dimensions $d$ where $d = 2(p+1)$; in other words, the rank of the current $p+1$ is half of the dimension. We will show that {\it conformal} field theories with such currents always have a free realization of the current in a sense that we will make precise. 

For notational convenience, define $n \equiv p+1 = \frac{d}{2}$ and consider the two-point function of two currents in a conformal theory on $\mathbb{R}^{d}$, which is completely fixed by scale invariance and conservation up to an overall prefactor $k$:
\begin{align}
& \langle J^{\mu_1 \cdots \mu_{n}}(x) J^{\rho_1 \cdots \rho_{n}}(0) \rangle = \nonumber \\
& \frac{k}{x^{d}}\le( g^{\mu_1 \rho_1} \cdots g^{\mu_{n} \rho_{n}} - d\frac{x^{\mu_1} x^{\rho_1}}{x^2} g^{\mu_2 \rho_2} \cdots g^{\mu_n \rho_n}\ri) \, . \label{2pointCFT}
\end{align} 
In this expression, it should be understood that the right-hand side is always antisymmetrized over both $\mu_i$ and $\rho_i$. There are only two possible tensor structures that can appear, and their relative factor has been fixed by demanding that $\p_{\mu_1} J^{\mu_1 \cdots \mu_{n}}(x) = 0$. 

Let us now define the ``dual'' current $J_{\star}$ as:
\be
J_{\star} \equiv \star J \, , \qquad J_{\star}^{\nu_1 \cdots \nu_n} \equiv \frac{1}{n!} \ep^{\nu_1 \cdots \nu_n}_{\phantom{\nu_1 \cdots \nu_n} \mu_1 \cdots \mu_n} J^{\mu_1 \cdots \mu_n} \, .
\ee
We now compute the two-point function of the divergence of $J_{\star}$ with $J$. From \eqref{2pointCFT} it is a few lines of algebra to show that
\be
\p_{\nu_1} \langle J^{\nu_1 \cdots \nu_{n}}_{\star}(x) J^{\rho_1 \cdots \rho_{n}}(0) \rangle = 0 \, ,
\ee
where the equality depends on the precise form of \eqref{2pointCFT} and will not generically hold in a non-scale-invariant theory. This implies that
\be
 \langle \p_{\nu_1} J^{\nu_1 \cdots \nu_{n}}_{\star}(x) \p_{\rho_1} J_{\star}^{\rho_1 \cdots \rho_{n}}(0) \rangle = 0 \, .
 \ee
Thus the operator $\p_{\nu_1} J_{\star}^{\nu_1 \cdots \nu_{n}}$ has vanishing two-point function in the vacuum. However, this two-point function measures the norm of the state created by acting with the operator $\p_{\nu_1} J_{\star}^{\nu_1 \cdots \nu_{n}}$ on the vacuum. This norm must be positive in a unitary CFT; it can only vanish if the operator is itself zero. We conclude that just like $J$, its Hodge dual $J_{\star}$ is also a divergenceless current. 

Using the fact that $\star^2 = \pm 1$, we thus find that we have
\be
d\star J = 0 \, , \qquad d J = 0 \, .
\ee
These two equations together mean that we can always locally write $J$ in terms of a $\frac{d}{2}-1$-form $B$ that obeys a free wave equation:
\be
J = dB \, ,\qquad  d\star dB = 0 \, .
\ee
Thus the correlation functions of the current can always be obtained from a free theory. This free theory is nothing else than the theory of Goldstone modes discussed above (if $p>0$). Therefore $CFT$s in this class are always in a \textit{spontaneously broken} phase. This can only happen for $d=2(p+1)$ as only then, is the stiffness (\ref{stiff}) a dimensionless parameter. Conversely, also only when $d=2(p+1)$, a symmetry \textit{broken} phase always corresponds to a $CFT.$ While free, these theories are still parameterized by the dimensionless stiffness and the spectrum of (non-local) operators depends on it.

Let us now discuss the implications of this result. Consider first the case $d=2$, $p=0$. Then we have derived the well-known fact in $CFT_2$ that if a vector current $J$ is conserved, its axial counterpart $\star_2 J$ is also conserved. This allows us to break the current into separately conserved holomorphic and antiholomorphic pieces and is the starting point for the construction of the $U(1)$ Kac-Moody algebra. $B$ is a free compact scalar, and the representation of the current algebra in terms of this free scalar is normally called bosonization. 

We now turn to $d =4$, $p=1$. It is helpful to think of the case of electromagnetism: if we have a conserved $2$-form current $J$ for magnetic flux, conformality implies that we {\it also} have a conserved $2$-form current for {\it electric} flux. In other words, we cannot have electric charges. The theory is necessarily free 4d electrodynamics with $B$ the free (magnetic) photon: we have shown that any 1-form current can be \textit{photonized}. 

Normally the logic here is reversed, and one says that the presence of electric charges makes the $U(1)$ coupling run logarithmically, spoiling conformality. The argument here shows that this result is more general than the perturbative context in which it usually appears: i.e. there is no way to have a conformal field theory with both a conserved magnetic flux and with dynamical electric charges. Indeed this was recently demonstrated in a holographic example \cite{Hofman:2017vwr} where non-conservation of $\star J$ was correlated with a lack of conformality. See also \cite{Grozdanov:2017kyl, Grozdanov:2018ewh} for related work.


\section{Abelian Kac-Moody algebra for 1-form symmetries in Maxwell theory}
In this section we specialize to a 4d CFT with a 1-form current $J$ and show that we may construct infinitely many conserved charges. These charges will be constructed on codimension-$1$ manifolds, and so, unlike higher-form charges \cite{Gaiotto:2014kfa}, are allowed to have nontrivial commutators. We will show that they form an algebra with central extension that can be thought of as a higher-form generalization of the Abelian Kac-Moody algebra in 2d CFT. In this section, we will compute commutators in the free Maxwell theory and in the next we will show how they can be obtained generically in any $CFT_4$ showing explicitly that they can always be photonized as a consequence of the conservation equation and conformal invariance.

We begin by constructing the following projectors on $2$-forms in Lorentzian $(3+1)$d:
\bea
P = \ha\le(1 + i \star_4\ri) \, , &\quad&  \bar P = \ha\le(1 - i \star_4\ri)\, , \label{projectors}\\
PA_2 \wedge B_2 = A_2 \wedge P B_2 \, ,&\quad& \bar P A_2 \wedge B_2 = A_2 \wedge \bar P B_2 \, ,\quad\quad
\eea
where the second equality shows how to pass the projector through a product of 2-forms. As shown previously, in a $CFT_4$ $J$ and $\star_4 J$ are both necessarily conserved, and thus the current may be split into separately conserved self-dual and anti-self-dual currents $j$ and $\bar j$. 
\be
j \equiv P J \, ,\quad \bar j \equiv \bar P J \, ,\quad d\star_4 j = d\star_4 \bar j =  d j = d \bar j = 0 \, .
\ee
These are the analogues of left and right-moving currents in 2d CFT. Specializing to the anti self-dual sector, consider now a \textit{chiral} 1-form $\Lam$ that satisfies
\be
\bar P \, d\Lam = 0 \label{chiral} \, .
\ee
For {\it any} choice of such a 1-form, the $3$-form $\bar j \, \wedge \,  \Lambda$ is closed: 
\be
d(\bar j \, \wedge \, \Lam) =\bar j \, \wedge \, d \Lam  = \bar j \, \wedge \,P\, d \Lam =  P\, \bar j \, \wedge \, d \Lam = 0\, .
\ee
Thus we can integrate this form over a 3-manifold to construct a conserved charge $\bar Q(\Lam)$.
\be
\bar Q( \Lam) =\int \bar j \, \wedge \, \Lam \label{charge}\, .
\ee 
We may do the same in the self-dual sector. We see that we have infinitely many conserved charges $\bar Q(\Lam)$ and $Q(\bar \Lam)$. Note that shifting $\Lam \to \Lam + d\phi$ with $\phi$ a $0$-form does not alter the charges (up to boundary terms); thus there is a gauge redundancy in the parametrization of charges. The same holds for $\bar \Lam$.

Above we have argued that any theory with this symmetry structure can be photonized to free electrodynamics, and so it is thus perhaps not surprising that a free theory has infinitely many conserved charges. They should be thought of as higher-form analogues of the infinitely many left and right moving charges of an abelian current in 2d CFT. In particular, in the 2d case the analogue of the chirality condition \eqref{chiral} is simply a restriction to holomorphic and anti-holomorphic test functions. 

It is also intriguing to note that though we started with a 2-form current,  the charges $\bar Q(\Lam)$ and $Q(\bar \Lam)$ are defined by integrals on 3-dimensional manifolds and thus are similar to ``ordinary'' charges arising from a 0-form symmetry. In particular, unlike charges defined by integrals on lower-dimensional manifolds \cite{Gaiotto:2014kfa}, any two of these charges will have a definite ordering in spacetime and thus can have non-trivial commutators. 

We now compute these commutators. We will do this using free Maxwell theory in this section. In the following section we develop a formalism that can be used to perform this calculation without resorting to the free field realization and, thus, showing explicitly the phenomenon of \textit{photonization}. Consider
\be
S = -\frac{1}{4g^2} \int d^4x\,  F^2 \, ,\qquad J^{\mu\nu} \equiv \ha \ep^{\mu\nu\rho\sig} F_{\rho\sig} \, ,
\ee
with $F = dA$ and $A$ a photon field. With this normalization of the action $k$ as defined in \eqref{2pointCFT} satisfies $k = \frac{g^2}{2\pi^2}$. From the fundamental commutation relation 
\be
[A_i(x), F_{tj}(y)] = i g^2 \delta_{ij}\delta^{(3)}(x-y) 
\ee
we obtain the following commutator between the currents:
\be
[J^{ti}(x), J^{kl}(y)] = ig^2 \le(\delta^{ik} \frac{\p}{\p x^l} - \delta^{il} \frac{\p}{\p x^k}\ri) \delta^{(3)}(x-y) \ . 
\ee
It is then straightforward to derive the following mixed commutator between the charges, which we assume are constructed as integrals on a constant time-slice in flat 4d space:
\begin{equation}\label{comboring}
[\bar Q(\Lam), Q({\bar \Lam})]  = - \frac{g^2}{2} \int d^3x\;\ep^{ijk} \Lam_{i} \p_{j} \bar\Lam_k \, .
\end{equation}
The right-hand side is the advertised central extension. This can be seen to be a natural generalization to higher dimension of the Abelian Kac-Moody algebra in 2d, with the central term taking a very similar form and with the $U(1)$ gauge coupling playing the role of the Kac-Moody level\footnote{As we are studying an Abelian theory, the value of the gauge coupling has physical significance only once the normalization of periods of $F$ is fixed, e.g. by fixing the charge lattice for electric and magnetic charges.}. The diagonal commutators $[\bar Q({\Lam}), \bar Q({\Lam})]$ and $[Q({\bar \Lam}), Q({\bar\Lam})]$ are both total derivatives. This is an interesting difference from 2d, where it is instead the mixed commutator between holomorphic and anti-holomorphic currents that is trivial.


\section{Twistor languange for $CFT_4$'s}

In the case of $CFT_4$'s with a conserved 2-form the above discussion can be put in a completely covariant formalism in terms of twistor variables allowing for a computation without mention of the free field realization. This formalism puts these theories on equal footing to their $CFT_2$s cousins and allows for the use of complex analysis technology. We will show the phenomenon of \textit{photonization} in this language and develop a formalism to discuss charges covariantly, recovering the results from previous sections.

\subsection{Photonization}
 
It will prove useful to introduce spinorial notation. As usual, an $SO(4)$ vector index $\mu$ can be exchanged for a pair of $SU(2)$ spinorial indices $\alpha \dot\alpha$ with $\alpha, \dot \alpha = 1, 2$. Using this notation  we can write a generic antisymmetric tensor as
\be
J_{\mu \nu} \rightarrow J_{\alpha \dot{\alpha} \beta \dot{\beta}} = j_{\alpha \beta} \epsilon_{\dot{\alpha} \dot{\beta}} +  \bar j_{\dot\alpha \dot\beta} \epsilon_{\alpha \beta} \, ,
\ee
\noindent where $j$ and $\bar j$ are symmetric SU(2) tensors. Notice that $j$ and $\bar j$ are the self-dual and anti self-dual parts of the current 2-form.  In this notation $\epsilon$ acts like the metric tensor. For example, the square of the position vector is given by
\be
x^2 = \epsilon_{\alpha \beta} \epsilon_{\dot\alpha \dot\beta} x^{\alpha \dot\alpha} x^{\beta \dot\beta} \, .
\ee
Using this technology we can construct the most general form of two point functions between $j$ and $\bar j$ consistent with SU(2) symmetry. They are
\bea
\langle j^{\alpha \beta} j^{\gamma \delta} \rangle &=& f(x^2) \left(\epsilon^{\alpha \gamma} \epsilon^{\beta \delta} + \epsilon^{\beta \gamma} \epsilon^{\alpha \delta}\right)\label{jj1}\, ,\\
\langle j^{\alpha \beta} \bar j^{\dot \gamma \dot\delta} \rangle &=& g(x^2) \left(x^{\alpha \dot\gamma} x^{\beta \dot\delta} + x^{\beta \dot\gamma} x^{\alpha \dot\delta}\right)\label{jj2}\, ,\\
\langle \bar j^{\dot\alpha \dot\beta} \bar j^{\dot\gamma \dot\delta} \rangle &=& h(x^2) \left(\epsilon^{\dot\alpha \dot\gamma} \epsilon^{\dot\beta \dot\delta} + \epsilon^{\dot\beta \dot\gamma} \epsilon^{\dot\alpha \dot\delta}\right)\label{jj3} \, .
\eea
If we now demand conformal symmetry and use the fact that $j$ and $\bar j$ have weight 2 under rescalings we obtain:
\bea\label{confkill}
f(x^2) \sim \frac{1}{x^4} \, , \quad g(x^2) \sim \frac{1}{x^6}\, , \quad h(x^2) \sim \frac{1}{x^4}\, ,
\eea
\noindent up to multiplicative constants.

Let us now add the requirement that the currents are conserved, i.e.
\bea
\partial_{\alpha \dot \beta} \, j^{\alpha}_{\phantom{\alpha} \beta} +\partial_{\beta \dot \alpha} \, \bar j^{\dot\alpha}_{\phantom{\dot\alpha} \dot \beta} =0 \, .
\eea
This implies $f=h=0$. The only nonzero correlation is, therefore, of the form:
\be\label{currenttwi}
\langle j^{\alpha \beta} \bar j^{\dot \gamma \dot\delta} \rangle =\frac{k}{2} \frac{x^{\alpha \dot\gamma} x^{\beta \dot\delta} + x^{\beta \dot\gamma} x^{\alpha \dot\delta}}{x^6} \, .
\ee

The absence of null states makes the conservation equations stronger, as in $CFT_2$, implying separately the conservation of the dual and self-dual currents.

\bea
\partial_{\alpha \dot \beta} \, j^{\alpha}_{\phantom{\alpha} \beta} =0 \, ,\\
\partial_{\beta \dot \alpha} \, \bar j^{\dot\alpha}_{\phantom{\dot\alpha} \dot \beta} =0\, .
\eea

This is equivalent to the statement that currents can be represented by free Maxwell fields where the above equations relate directly to the conservation of electric and magnetic fluxes. Therefore our $CFT_4$ has photonized

\subsection{Infinite dimensional charge algebra}

At this point it pays off to use twistor technology \cite{Wolf:2010av} to write down the currents in terms of operators that are holomorphic in $\mathbb{CP}^3$, parameterized by twistorial variables $(z^\alpha, \lambda_{\dot\alpha})$. We write:
\bea
j_{\alpha \beta}(x) &=& \frac{1}{2\pi i} \oint d\lambda^{\dot{\omega}} \lambda_{\dot{\omega}}  \, \frac{\partial}{\partial z^\alpha} \frac{\partial}{\partial z^\beta} \, Q(z, \lambda)\label{jtwist} \, ,\\
\bar j_{\dot\alpha \dot\beta}(x) &=& \frac{1}{2\pi i}  \oint d\lambda^{\dot{\omega}} \lambda_{\dot{\omega}} \, \lambda_{\dot{\alpha}}  \lambda_{\dot{\beta}} \, \bar Q(z, \lambda)\label{jbartwist}\, ,
\eea
\noindent where the integrals are over a closed contour in a $\mathbb{CP}^1$ parameterized by $(z^\alpha, \lambda_{\dot\alpha})=(x^{\alpha \dot \alpha} \lambda_{\dot{\alpha}}, \lambda_{\dot{\alpha}})$. The $\lambda_{\dot\omega}$ integrals are always understood at fixed $x^{\alpha \dot \alpha}$, not $z^\alpha$.

With this notation it is straight forward to check that the following correlator for the $Q$'s reproduces (\ref{currenttwi}).
\be
\big\langle \, Q(z, \lambda) \, \bar Q(z', \lambda') \, \big\rangle = \pi i \, k \frac{\epsilon^{\alpha \beta}B_\alpha C_\beta \, \delta([ \lambda \lambda' ] )}{B_\gamma (z-z')^\gamma C_\delta (z-z')^\delta} \frac{[\lambda \mu']}{[\lambda ' \mu']} \, ,
\ee
\noindent where $[ \lambda \lambda' ] =\epsilon^{\dot \alpha \dot \beta} \lambda_{\dot{\alpha}} \lambda'_{\dot{\beta}}$,  $\mu'$ is a non collinear spinor to $\lambda'$ and $B$ and $C$ are two independent spinors. Integration over the above expression over $\lambda'$ and $\lambda$ proves independent of $\mu'$, $B$ and $C$. A more covariant version of the above formula can be obtained by replacing  $\delta([ \lambda \lambda' ] )\frac{[\lambda \mu']}{[\lambda ' \mu']} \rightarrow \frac{1}{[\lambda \lambda']}$ and giving a contour prescription for the integrals in  (\ref{jtwist}) and (\ref{jbartwist})

The formula above is important as it puts correlators in terms of a complex functions that can be integrated by residues, exactly as in the more familiar $CFT_2$ context.

It turns out that the $Q$ operators have a clear physical interpretation in terms of usual charges integrated over compact codimension one surfaces. Consider as in the previous section a \textit{chiral} 1-form $\Lambda_{\alpha \dot{\alpha}}$ such that its exterior derivative is strictly self dual, $d\Lambda = (d\Lambda)_{\alpha\beta} \, \epsilon_{\dot{\alpha} \dot\beta}$. Then we can build conserved charges
\be
\bar Q(\Lambda) = \int \bar j \wedge \Lambda  \, .
\ee
Notice that charges defined on compact surfaces require that $\Lambda \neq d\phi$, as this would render them trivial as a consequence of the conservation of the 2-form currents.
Solving the constraints that $\Lambda$ satisfies is hard in physical space but straightforward in twistor space
\be\label{Atwist}
\Lambda_{\alpha \dot{\alpha}} = \frac{1}{2 \pi i} \oint d\lambda^{\dot{\omega}} \lambda_{\dot{\omega}}  \, \frac{\mu_{\dot \alpha}}{[\mu \lambda]}  \frac{\partial}{\partial z^\alpha} \varphi(z,\lambda) \, .
\ee
Using (\ref{jbartwist}) and (\ref{Atwist}) we can write, by going to momentum space $k_\alpha$ dual to $z^\alpha$,
\be
\bar Q(\Lambda) = \frac{1}{2 \pi i}  \oint d\lambda^{\dot{\omega}} \lambda_{\dot{\omega}}  \, \frac{d^2 k}{\left( 2 \pi\right)^2} \, \bar Q(k,\lambda) \varphi(-k,\lambda) \, .
\ee
If we pick a basis of functions:
\be
\varphi_{z, \lambda'}(k,\lambda) = e^{i k_\alpha z^\alpha}  2 \pi i \delta([\lambda' \lambda]) \,  \frac{[\lambda' \mu]}{[\lambda \mu]} 
\ee
and define
\be
\Lambda_{z,\lambda} = \frac{1}{2\pi i} \oint d\lambda'^{\dot{\omega}} \lambda'_{\dot{\omega}}  \, \frac{\mu_{\dot \alpha}}{[\mu \lambda']}  \frac{\partial}{\partial z'^\alpha} \int  \frac{d^2 k}{\left( 2 \pi\right)^2} \, e^{ -i k_\alpha z'^\alpha} \, \varphi_{z, \lambda}(k,\lambda') 
\ee

\noindent we obtain:
\be
\bar Q(\Lambda_{z,\lambda}) = \bar Q(z,\lambda) \, ,
\ee
\noindent which shows that the twistor space operators are nothing else than charges. The equivalent statement can be made for $Q(z,\lambda)$ using 1-forms $\bar \Lambda$ with anti-self dual $d \bar \Lambda$, 
\be
Q(\Lambda) = \int j \wedge \bar \Lambda \, .
\ee

Now we can exploit this language to derive a Kac-Moody algebra satisfied by the conserved charges $Q(\bar \Lambda)$ and $\bar Q(\Lambda)$. In this language the calculation is completely analogous to the residue computation familiar in $CFT_2$. Let us compute
\bea
\epsilon_{\dot \alpha \dot \beta}\big[  j_{\alpha \beta}(x) \, ,  \bar Q(\Lambda) \big] = \quad\quad \quad\quad\quad\quad & &\quad \nonumber\\
\frac{\epsilon_{\dot \alpha \dot \beta}}{2} \int \big\langle j_{\alpha \beta}(x) \,  \bar j_{\dot \gamma \dot \delta}(y) \big\rangle \, \epsilon_{\gamma \delta} \Lambda_{\rho \dot \rho} \, dy^{\gamma \dot \gamma} \wedge dy^{\delta \dot \delta} \wedge dy^{\rho \dot \rho}\, .\quad\quad 
\eea
The fact that the above computation represents a commutator is related to the choice of contours for the integrals as in $CFT_2$.
We can compute the above quantity by going as before to momentum space $k_\alpha$ and performing the space integrals. The result is:
\bea
\epsilon_{\dot \alpha \dot \beta} \big\langle  j_{\alpha \beta}(x) \, \bar Q(\Lambda) \big\rangle &=&   k\,  \frac{\pi  i}{2}\oint d\lambda^{\dot{\omega}} \lambda_{\dot{\omega}}  \,  \frac{\partial}{\partial z^\alpha} \frac{\partial}{\partial z^\beta} \varphi(z, \lambda) \frac{\mu_{[\dot{\alpha}} \lambda_{\dot{\beta}]}}{[\mu \lambda]}  \nonumber \\
 & = & k\, \frac{\pi i}{2}\frac{\partial}{\partial x^{\beta [ \dot\beta}} \oint d\lambda^{\dot{\omega}} \lambda_{\dot{\omega}}  \,   \frac{\mu_{\dot{\alpha}]}}{[\mu \lambda]} \frac{\partial}{\partial z^\alpha} \varphi(z, \lambda)\nonumber\\
 &=& - \pi^2 k \left(d\Lambda\right)_{\alpha \beta} \epsilon_{\dot \alpha \dot \beta} \, \quad .
\eea
We can now integrate this against a 1-form $\bar \Lam$ to obtain the charge commutator
\be
\big[ Q(\bar \Lambda), \bar Q (\Lambda) \big] =  -\pi^2 k \int d\Lambda \wedge \bar \Lambda \, ,
\ee

\noindent which agrees with (\ref{comboring}) and is consistent with the \textit{photonization} of $CFT_4$'s. 

We expect that this formalism will prove useful in understanding four dimensional gauge theories in a gauge invariant manner. \\
\

\section{Relation to soft photon theorems}
In non-conformal theories (or, equivalently, theories with dynamical electric charges) we cannot argue, as in (\ref{confkill}), that the self-dual and anti self-dual parts of the currents are conserved. Therefore the construction from the previous section is absent and there is no non-trivial algebra of charges for theories defined on compact manifolds. 

If the theories are quantized, however, over {\it non-compact} space-like surfaces $\Sigma$, we can mimic part of the discussion above. The charges are now parameterized by flat connections $\mathcal{A} = d \phi$ as:
\be
\mathcal{Q}(\mathcal{A}) = \int_\Sigma\ \star_4 J \wedge \mathcal{A} = \int_{\partial \Sigma} \star_4 J \wedge \phi
\ee

These charges are (up to electric-magnetic duality) precisely those discussed in the asymptotic symmetry approach to soft theorems \cite{Strominger:2013lka, He:2014cra} (see \cite{Strominger:2017zoo} for a review). We see that these charges arise in a fully gauge-invariant formalism and they extend to any QFT enjoying higher form symmetries. We thus expect that the Kac-Moody algebra appearing in \cite{Strominger:2013lka} can be reproduced from \eqref{jj1}-\eqref{jj3}. It would be interesting to better understand this connection.

\section{Conclusions}

We conclude this short note with some directions for further research.  Our proof for the existence of Goldstone modes made heavy use of Lorentz invariance. We do not believe this is necessary and it would be very interesting to relax this assumption. This has direct relevance to lattice models of emergent gauge fields and finite-temperature dynamics. 

Furthermore, it would be of great interest to extend the results from this work to non-abelian gauge theories and gravity. While there are known obstructions to the construction of higher form currents for non-abelian algebras \cite{Gaiotto:2014kfa}, it is tempting to interpret massless degrees of freedom in these theories (when they are free in the IR) in terms of corresponding Goldstone modes.

We also find the similarity of the twistor formalism for $CFT_4$'s to the complex analysis tools in $CFT_2$'s to be intriguing.  This suggests that (at least some sectors of) some higher dimensional $CFT$'s might show as much structure as the well studied two dimensional case\footnote{See \cite{Beem:2013sza,Beem:2014kka} for examples of this phenomenon in the presence of supersymmetry and also \cite{Losev:1995cr} for examples including a WZW term.}. We have only begun to work out the details of the twistorial approach to higher form charges. In particular, 4d conformal invariance acts as an $SU(4)$ in twistor space, which we did not attempt to keep manifest above. Another advantage of the twistorial approach is that is provides a natural language to discuss the physics of light-ray operators which has proven of fundamental importance to understand the UV consistency conditions on the space of $CFT$'s \cite{Hofman:2008ar, Hofman:2009ug, Faulkner:2016mzt, Hartman:2016lgu}

Lastly, let us add that while we have not worked out the details, we also expect a similar structure for massless 2-form fields in 6d, which could have implications towards the understanding of the $(2,0)$ $CFT_6$. See \cite{Afshar:2018apx} for an interesting discussion of asymptotic symmetries in p-form theories.


Conceptually, this work can be viewed as progress towards organizing quantum field theory in terms of the dynamics of extended objects (which are charged under higher form currents and are gauge invariant) rather than local operators (which are charged under ordinary currents and are typically not gauge invariant)\footnote{See \cite{Friedan:2016mvo,Friedan:2017yer} for different attempts in this direction.}. We anticipate further developments -- both applied and formal -- from this program. 
 
{\bf Note added}: In the final stages of preparation of this paper, two pre-prints appeared on the arXiv with some overlap with this work: \cite{Cordova:2018cvg} has some overlap with Section \ref{sec:photonization}, and \cite{Lake:2018dqm} presents an alternative route to the gapless Goldstone modes in Section \ref{sec:goldstone}.

\vspace{0.2in}   \centerline{\bf{Acknowledgements}} \vspace{0.2in} We thank S. Cremonesi, T. Griffin, E. Katz, S. Melville, S. Ross, S. Vandoren and D. Woldram for illuminating discussions. NI would like to thank Delta ITP at the University of Amsterdam for hospitality during the completion of this work. NI is supported in part by the STFC under consolidated grant ST/L000407/1. This work is part of the Delta ITP consortium, a program of the NWO that is funded by the Dutch Ministry of Education, Culture and Science (OCW). This project has received funding from the European Research Council (ERC) under the European Union’s Horizon 2020 research and innovation programme (grant agreement No 715656).

\bibliographystyle{utphys}
\bibliography{all}

\providecommand{\href}[2]{#2}\begingroup\raggedright\begin{thebibliography}{10}

\bibitem{Gaiotto:2014kfa}
D.~Gaiotto, A.~Kapustin, N.~Seiberg, and B.~Willett, ``{Generalized Global
  Symmetries},'' \href{http://dx.doi.org/10.1007/JHEP02(2015)172}{{\em JHEP}
  {\bfseries 02} (2015) 172},
\href{http://arxiv.org/abs/1412.5148}{{\ttfamily arXiv:1412.5148 [hep-th]}}.

\bibitem{Kapustin:2005py}
A.~Kapustin, ``{Wilson-'t Hooft operators in four-dimensional gauge theories
  and S-duality},'' \href{http://dx.doi.org/10.1103/PhysRevD.74.025005}{{\em
  Phys. Rev.} {\bfseries D74} (2006) 025005},
\href{http://arxiv.org/abs/hep-th/0501015}{{\ttfamily arXiv:hep-th/0501015
  [hep-th]}}.

\bibitem{Aharony:2013hda}
O.~Aharony, N.~Seiberg, and Y.~Tachikawa, ``{Reading between the lines of
  four-dimensional gauge theories},''
  \href{http://dx.doi.org/10.1007/JHEP08(2013)115}{{\em JHEP} {\bfseries 08}
  (2013) 115},
\href{http://arxiv.org/abs/1305.0318}{{\ttfamily arXiv:1305.0318 [hep-th]}}.

\bibitem{Gaiotto:2017yup}
D.~Gaiotto, A.~Kapustin, Z.~Komargodski, and N.~Seiberg, ``{Theta, Time
  Reversal, and Temperature},''
  \href{http://dx.doi.org/10.1007/JHEP05(2017)091}{{\em JHEP} {\bfseries 05}
  (2017) 091},
\href{http://arxiv.org/abs/1703.00501}{{\ttfamily arXiv:1703.00501 [hep-th]}}.

\bibitem{Komargodski:2017dmc}
Z.~Komargodski, A.~Sharon, R.~Thorngren, and X.~Zhou, ``{Comments on Abelian
  Higgs Models and Persistent Order},''
\href{http://arxiv.org/abs/1705.04786}{{\ttfamily arXiv:1705.04786 [hep-th]}}.

\bibitem{PhysRevB.93.155131}
B.~Yoshida, ``Topological phases with generalized global symmetries,''
  \href{http://dx.doi.org/10.1103/PhysRevB.93.155131}{{\em Phys. Rev. B}
  {\bfseries 93} (Apr, 2016) 155131}.
  \url{https://link.aps.org/doi/10.1103/PhysRevB.93.155131}.

\bibitem{Wang:2014pma}
J.~C. Wang, Z.-C. Gu, and X.-G. Wen, ``{Field theory representation of
  gauge-gravity symmetry-protected topological invariants, group cohomology and
  beyond},'' \href{http://dx.doi.org/10.1103/PhysRevLett.114.031601}{{\em Phys.
  Rev. Lett.} {\bfseries 114} no.~3, (2015) 031601},
\href{http://arxiv.org/abs/1405.7689}{{\ttfamily arXiv:1405.7689
  [cond-mat.str-el]}}.

\bibitem{Tanizaki:2017mtm}
Y.~Tanizaki, Y.~Kikuchi, T.~Misumi, and N.~Sakai, ``{Anomaly matching for phase
  diagram of massless $\mathbb{Z}_N$-QCD},''
\href{http://arxiv.org/abs/1711.10487}{{\ttfamily arXiv:1711.10487 [hep-th]}}.

\bibitem{Kitano:2017jng}
R.~Kitano, T.~Suyama, and N.~Yamada, ``{$\theta=\pi$ in $SU(N)/\mathbb{Z}_N$
  gauge theories},'' \href{http://dx.doi.org/10.1007/JHEP09(2017)137}{{\em
  JHEP} {\bfseries 09} (2017) 137},
\href{http://arxiv.org/abs/1709.04225}{{\ttfamily arXiv:1709.04225 [hep-th]}}.

\bibitem{Gaiotto:2017tne}
D.~Gaiotto, Z.~Komargodski, and N.~Seiberg, ``{Time-reversal breaking in
  QCD$_{4}$, walls, and dualities in 2 + 1 dimensions},''
  \href{http://dx.doi.org/10.1007/JHEP01(2018)110}{{\em JHEP} {\bfseries 01}
  (2018) 110},
\href{http://arxiv.org/abs/1708.06806}{{\ttfamily arXiv:1708.06806 [hep-th]}}.

\bibitem{Grozdanov:2016tdf}
S.~Grozdanov, D.~M. Hofman, and N.~Iqbal, ``{Generalized global symmetries and
  dissipative magnetohydrodynamics},''
  \href{http://dx.doi.org/10.1103/PhysRevD.95.096003}{{\em Phys. Rev.}
  {\bfseries D95} no.~9, (2017) 096003},
\href{http://arxiv.org/abs/1610.07392}{{\ttfamily arXiv:1610.07392 [hep-th]}}.

\bibitem{Levin:2004mi}
M.~A. Levin and X.-G. Wen, ``{String net condensation: A Physical mechanism for
  topological phases},''
  \href{http://dx.doi.org/10.1103/PhysRevB.71.045110}{{\em Phys. Rev.}
  {\bfseries B71} (2005) 045110},
\href{http://arxiv.org/abs/cond-mat/0404617}{{\ttfamily arXiv:cond-mat/0404617
  [cond-mat]}}.

\bibitem{hansson2004superconductors}
T.~Hansson, V.~Oganesyan, and S.~Sondhi, ``Superconductors are topologically
  ordered,'' {\em Annals of Physics} {\bfseries 313} no.~2, (2004) 497--538.

\bibitem{Rosenstein:1990py}
B.~Rosenstein and A.~Kovner, ``{Masslessness of photon and Goldstone
  theorem},''
\href{http://dx.doi.org/10.1142/S0217751X91001726}{{\em Int. J. Mod. Phys.}
  {\bfseries A6} (1991) 3559--3570}.

\bibitem{Kovner:1990pz}
A.~Kovner, B.~Rosenstein, and D.~Eliezer, ``{Photon as a Goldstone boson in
  (2+1)-dimensional Abelian gauge theories},''
\href{http://dx.doi.org/10.1016/0550-3213(91)90263-W}{{\em Nucl. Phys.}
  {\bfseries B350} (1991) 325--354}.

\bibitem{Hofman:2017vwr}
D.~M. Hofman and N.~Iqbal, ``{Generalized global symmetries and holography},''
  \href{http://dx.doi.org/10.21468/SciPostPhys.4.1.005}{{\em SciPost Phys.}
  {\bfseries 4} (2018) 005},
\href{http://arxiv.org/abs/1707.08577}{{\ttfamily arXiv:1707.08577 [hep-th]}}.

\bibitem{Grozdanov:2017kyl}
S.~Grozdanov and N.~Poovuttikul, ``{Generalised global symmetries and
  magnetohydrodynamic waves in a strongly interacting holographic plasma},''
\href{http://arxiv.org/abs/1707.04182}{{\ttfamily arXiv:1707.04182 [hep-th]}}.

\bibitem{Grozdanov:2018ewh}
S.~Grozdanov and N.~Poovuttikul, ``{Generalized global symmetries in states
  with dynamical defects: The case of the transverse sound in field theory and
  holography},'' \href{http://dx.doi.org/10.1103/PhysRevD.97.106005}{{\em Phys.
  Rev.} {\bfseries D97} no.~10, (2018) 106005},
\href{http://arxiv.org/abs/1801.03199}{{\ttfamily arXiv:1801.03199 [hep-th]}}.

\bibitem{Wolf:2010av}
M.~Wolf, ``{A First Course on Twistors, Integrability and Gluon Scattering
  Amplitudes},'' \href{http://dx.doi.org/10.1088/1751-8113/43/39/393001}{{\em
  J. Phys.} {\bfseries A43} (2010) 393001},
\href{http://arxiv.org/abs/1001.3871}{{\ttfamily arXiv:1001.3871 [hep-th]}}.

\bibitem{Strominger:2013lka}
A.~Strominger, ``{Asymptotic Symmetries of Yang-Mills Theory},''
  \href{http://dx.doi.org/10.1007/JHEP07(2014)151}{{\em JHEP} {\bfseries 07}
  (2014) 151},
\href{http://arxiv.org/abs/1308.0589}{{\ttfamily arXiv:1308.0589 [hep-th]}}.

\bibitem{He:2014cra}
T.~He, P.~Mitra, A.~P. Porfyriadis, and A.~Strominger, ``{New Symmetries of
  Massless QED},'' \href{http://dx.doi.org/10.1007/JHEP10(2014)112}{{\em JHEP}
  {\bfseries 10} (2014) 112},
\href{http://arxiv.org/abs/1407.3789}{{\ttfamily arXiv:1407.3789 [hep-th]}}.

\bibitem{Strominger:2017zoo}
A.~Strominger, ``{Lectures on the Infrared Structure of Gravity and Gauge
  Theory},''
\href{http://arxiv.org/abs/1703.05448}{{\ttfamily arXiv:1703.05448 [hep-th]}}.

\bibitem{Beem:2013sza}
C.~Beem, M.~Lemos, P.~Liendo, W.~Peelaers, L.~Rastelli, and B.~C. van Rees,
  ``{Infinite Chiral Symmetry in Four Dimensions},''
  \href{http://dx.doi.org/10.1007/s00220-014-2272-x}{{\em Commun. Math. Phys.}
  {\bfseries 336} no.~3, (2015) 1359--1433},
\href{http://arxiv.org/abs/1312.5344}{{\ttfamily arXiv:1312.5344 [hep-th]}}.

\bibitem{Beem:2014kka}
C.~Beem, L.~Rastelli, and B.~C. van Rees, ``{$ \mathcal{W} $ symmetry in six
  dimensions},'' \href{http://dx.doi.org/10.1007/JHEP05(2015)017}{{\em JHEP}
  {\bfseries 05} (2015) 017},
\href{http://arxiv.org/abs/1404.1079}{{\ttfamily arXiv:1404.1079 [hep-th]}}.

\bibitem{Losev:1995cr}
A.~Losev, G.~W. Moore, N.~Nekrasov, and S.~Shatashvili, ``{Four-dimensional
  avatars of two-dimensional RCFT},''
  \href{http://dx.doi.org/10.1016/0920-5632(96)00015-1}{{\em Nucl. Phys. Proc.
  Suppl.} {\bfseries 46} (1996) 130--145},
  \href{http://arxiv.org/abs/hep-th/9509151}{{\ttfamily arXiv:hep-th/9509151
  [hep-th]}}.
[,336(1995)].

\bibitem{Hofman:2008ar}
D.~M. Hofman and J.~Maldacena, ``{Conformal collider physics: Energy and charge
  correlations},'' \href{http://dx.doi.org/10.1088/1126-6708/2008/05/012}{{\em
  JHEP} {\bfseries 05} (2008) 012},
\href{http://arxiv.org/abs/0803.1467}{{\ttfamily arXiv:0803.1467 [hep-th]}}.

\bibitem{Hofman:2009ug}
D.~M. Hofman, ``{Higher Derivative Gravity, Causality and Positivity of Energy
  in a UV complete QFT},''
  \href{http://dx.doi.org/10.1016/j.nuclphysb.2009.08.001}{{\em Nucl. Phys.}
  {\bfseries B823} (2009) 174--194},
\href{http://arxiv.org/abs/0907.1625}{{\ttfamily arXiv:0907.1625 [hep-th]}}.

\bibitem{Faulkner:2016mzt}
T.~Faulkner, R.~G. Leigh, O.~Parrikar, and H.~Wang, ``{Modular Hamiltonians for
  Deformed Half-Spaces and the Averaged Null Energy Condition},''
  \href{http://dx.doi.org/10.1007/JHEP09(2016)038}{{\em JHEP} {\bfseries 09}
  (2016) 038},
\href{http://arxiv.org/abs/1605.08072}{{\ttfamily arXiv:1605.08072 [hep-th]}}.

\bibitem{Hartman:2016lgu}
T.~Hartman, S.~Kundu, and A.~Tajdini, ``{Averaged Null Energy Condition from
  Causality},'' \href{http://dx.doi.org/10.1007/JHEP07(2017)066}{{\em JHEP}
  {\bfseries 07} (2017) 066},
\href{http://arxiv.org/abs/1610.05308}{{\ttfamily arXiv:1610.05308 [hep-th]}}.

\bibitem{Afshar:2018apx}
H.~Afshar, E.~Esmaeili, and M.~M. Sheikh-Jabbari, ``{Asymptotic Symmetries in
  $p$-Form Theories},'' \href{http://dx.doi.org/10.1007/JHEP05(2018)042}{{\em
  JHEP} {\bfseries 05} (2018) 042},
\href{http://arxiv.org/abs/1801.07752}{{\ttfamily arXiv:1801.07752 [hep-th]}}.

\bibitem{Friedan:2016mvo}
D.~Friedan, ``{Quantum field theories of extended objects},''
\href{http://arxiv.org/abs/1605.03279}{{\ttfamily arXiv:1605.03279 [hep-th]}}.

\bibitem{Friedan:2017yer}
D.~Friedan, ``{A new kind of quantum field theory of (n-1)-dimensional defects
  in 2n dimensions},''
\href{http://arxiv.org/abs/1711.05049}{{\ttfamily arXiv:1711.05049 [hep-th]}}.

\bibitem{Cordova:2018cvg}
C.~Cordova, T.~T. Dumitrescu, and K.~Intriligator, ``{Exploring 2-Group Global
  Symmetries},''
\href{http://arxiv.org/abs/1802.04790}{{\ttfamily arXiv:1802.04790 [hep-th]}}.

\bibitem{Lake:2018dqm}
E.~Lake, ``{Higher-form symmetries and spontaneous symmetry breaking},''
\href{http://arxiv.org/abs/1802.07747}{{\ttfamily arXiv:1802.07747 [hep-th]}}.

\end{thebibliography}\endgroup

\end{document}